\documentclass[fleqn,twoside]{article}
\usepackage{espcrc2}
\usepackage{units}
\usepackage{upgreek}
\usepackage{graphicx}
\usepackage[figuresright]{rotating}

\title{A Scintillating Fiber Tracker With SiPM Readout}

\author{
	G. Roper Yearwood\address[RWTH]{I. Physikalisches Institut B, RWTH Aachen University, 52074 Aachen, Germany},
	B. Beischer\addressmark[RWTH],
	Ch.-H. Chung\addressmark[RWTH],
	Ph. v. Doetinchem\addressmark[RWTH],
	H. Gast\addressmark[RWTH],
	R. Greim\addressmark[RWTH],
	T. \nolinebreak Kirn\addressmark[RWTH],
	S. Schael\addressmark[RWTH],
	N. Zimmermann\addressmark[RWTH],
	T. Nakada\address[LAUSANNE]{Ecole Polytechnique F\'ed\'erale de Lausanne, Dorigny, CH-1015 Lausanne, Switzerland},
	G. Ambrosi\address[PERUGIA]{Istituto Nazionale di Fisica Nucleare, Sezione di Perugia, 06123 Perugia, Italy},
	P. Azzarello\addressmark[PERUGIA],
	R. Battiston\addressmark[PERUGIA]\address[PERUGIA2]{Universit\`a degli Studi di Perugia, Dipt. di Fisica, 06123 Perugia, Italy},
	C. \nolinebreak Piemonte\address[FBK]{Fondazione Bruno Kessler - Istituto per la Ricerca Scientifica e tecnologica, 38050 Trento, Italy}
}

\begin{document}

\begin{abstract}
We present a prototype for the first tracking detector consisting of $\unit[250]{\upmu m}$ thin scintillating fibers and silicon photomultiplier (SiPM) arrays. The detector has a modular design, each module consists of a mechanical support structure of $\unit[10]{mm}$ Rohacell foam between two $\unit[100]{\upmu m}$ thin carbon fiber skins. Five layers of scintillating fibers are glued to both top and bottom of the support structure. SiPM arrays with a channel pitch of $\unit[250]{\upmu m}$ are placed in front of the fibers.

We show the results of the first module prototype using multiclad fibers of types Bicron BCF-20 and Kuraray SCSF-81M that were read out by novel 32-channel SiPM arrays from FBK-irst/INFN Perugia as well as 32-channel SiPM arrays produced by Hamamatsu. A spatial resolution of $\unit[88]{\upmu m} \pm \unit[6]{\upmu m}$ at an average yield of 10 detected photons per minimal ionizig particle has been achieved.
\end{abstract}

\maketitle

\section{Tracker Module Prototypes}

\begin{figure}
\centering
\includegraphics[width=\columnwidth]{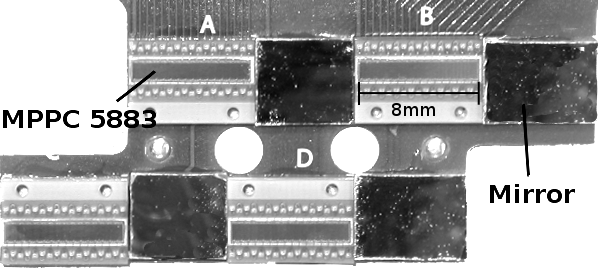}
\caption{One optical hybrid with 4 Hamamatsu MPPC 5883 SiPM arrays.}
\label{hpo_hamamatsu}
\end{figure}

A design for a modular charged-particle tracking-detector made up from scintillating fibers with silicon photomultiplier (SiPM) readout has been proposed for the PEBS balloon-borne spectrometer\cite{gast07}. Each module is built around a mechanical support from a $\unit[10]{mm}$ thick Rohacell foam layer contained between two $\unit[100]{\upmu m}$ thin carbon fiber skins. Two ribbons made from round multi-clad scintillating fibers with a nominal diameter of $\unit[250]{\upmu m}$ are glued to the mechanical support, one onto the top and one onto the bottom. Each ribbon consists of 5 layers of fibers. One layer consists of approximately $128$ fibers which are placed with a nominal pitch of $\unit[275]{\upmu m}$. The fiber layers of a ribbon are staggered by half of the fiber pitch. The fiber ribbons are glued with a Stycast epoxy loaded with $\unit[20]{\%}-\unit[33]{\%} \mathrm{TiO_2}$ to prevent optical crosstalk between the fibers. Optical readout hybrids (Fig.\ref{hpo_hamamatsu}) consisting of four linear SiPM arrays on a PCB board are placed on top of the polished fiber ends on both sides of the fiber module. Each fiber is read out by a SiPM array on one end and either covered by a mirror on the other side. 

\subsection{Silicon Photomultiplier Arrays}

\begin{table*}
\caption{Tested Silicon Photomultiplier Arrays.}
\begin{tabular*}{\textwidth}{@{\extracolsep\fill}lrrrrrr}
\hline
Type & PDE & $\lambda_{\mbox{\tiny{peak}}}/\unit{nm}$ & DCR $/\unit{kHz}$ & act. area$/\unit{mm}^2$ & no. channels & $U_{\mbox{\tiny{bias}}}/\unit{V}$\\
\hline 
Hamamatsu MPPC 5883 &  $50 \%$ & 450 & $200$ & $8.0 \times 1.1$ & $32$ & $71.0$\\
FBK-irst 2007 & $25 \%$ & $500$ & 550 & $8.0 \times 1.1$ & $32$ & $32.5$\\
\hline
\end{tabular*}
\label{arrays}
\end{table*}

Two types of SiPM arrays were available for testing (Tab.\ref{arrays}). The Hamamatsu MPPC 5883\cite{hamamatsu} achieves a higher photon detection efficiency (PDE) at a lower dark count rate (DCR) compared to the FBK-irst 2007\cite{piemonte06,piemonte07} due to its reduced number of pixels (80 pixels per channel for the MPPC 5883 compared to 110 pixels for the FBK-irst 2007) and a subsequently smaller relative dead area within the active area due to metallizations and quenching resistors\cite{greim08}. The MPPC 5883 is covered by a $\unit[275]{\upmu m}$ thick layer of optical glue. The FBK-irst 2007 comes without protective glue layer on top.

\subsection{Module Production and Precision}

\begin{table}[t]
\caption{Tested Scintillating Fibers.}
\begin{tabular*}{\columnwidth}{@{\extracolsep\fill}lrr}
\hline
Type & $\lambda_{\mbox{\tiny{peak}}}/\unit{nm}$ & $\oslash/\unit{mm}$\\
\hline 
SCSF-81M\cite{kuraray} &  $437$ & $0.230 \pm 0.011$\\
BCF-20\cite{bicron} & $492$ & $0.241 \pm 0.014$\\
\hline
\end{tabular*}
\label{fibers}
\end{table}

Two different types of fibers have been used during the production of the prototype modules (Tab.\ref{fibers}). The thickness of the scintillating fibers is measured optically during the first step of the module production. The tested Bicron BCF-20 fibers are elliptic with an eccentricity of approximately $\varepsilon = 0.5$ while the Kuraray SCSF-81M fibers are round. The measured tolerances for the fiber diameter are approximately $15\%$ of the mean diameter for both types of fibers.

Five layers of fibers are wound onto a drum with a helical groove using a CNC winding machine. The drum has a diameter of $\unit[200]{mm}$ and the grooves have a pitch of $\unit[0.275]{mm}$. Following the completion of each layer, an adhesive is applied to the fiber layer. After the adhesive has cured, the five fiber layers are cut and straightened out in order to obtain a $\unit[600]{mm}$ long fiber ribbon. Two ribbons are glued to the module support and the fiber ends are polished. The precision of the fiber placement is determined using a regular image scanner to take a picture of the fiber ends. The SCSF-81M fiber ribbons show the better mechanical precision. The fibers within one layer are placed with a pitch of $\unit[275]{\upmu m}$ and a precision (rms) of $\unit[19]{\upmu m}$. For the BCF-20 fiber ribbons a pitch of $\unit[273]{\upmu m}$ and a precision of $\unit[37]{\upmu m}$ were measured. The worse precision can be explained by the fact that the thickness of the BCF-20 fiber frequently exceeds the groove pitch of the drum used to wind the fiber ribbon thus forcing other fibers within the ribbon into disarray.

\section{Beamtest 2008}

\begin{figure}[t]
\centering
\includegraphics[width=\columnwidth]{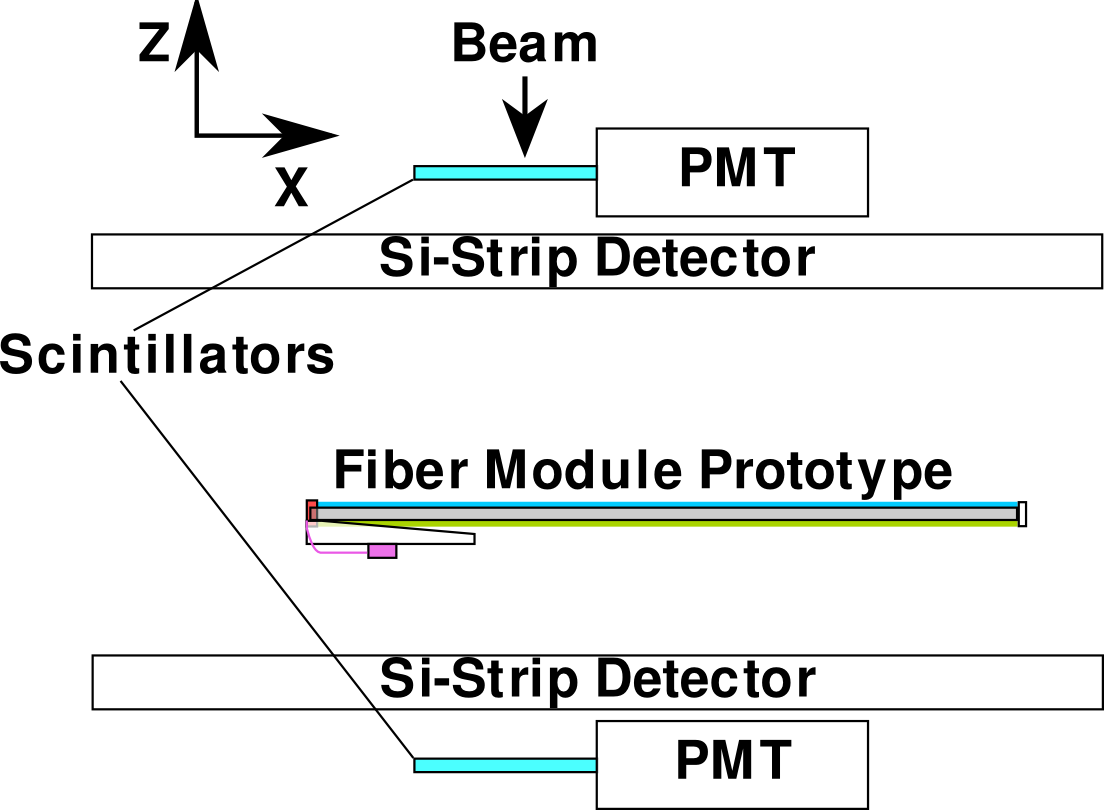}
\caption{The test setup used during the beamtest 2008.}
\label{beamtestsetup}
\end{figure}

The first module prototype was tested with a $\unit[10]{GeV}$ secondary proton/pion beam at the T9 beam line of the PS accelerator at CERN. The testbeam setup (Fig.\ref{beamtestsetup}) consisted of two trigger counters, two spare double-sided silicon strip detectors from the AMS silicon tracker project\cite{burger02} as a beam telescope and the fiber module prototype. The fiber module consisted of one SCSF-81M fiber ribbon and one BCF-20 fiber ribbon. Two optical hybrids were used during the beamtest to read out the fiber module prototype, one with four FBK-irst 2007 and another one with four MPPC 5883s. The  MPPC 5883s were coupled directly to the fibers, for the FBK-irst 2007 a nominal air gap of $\unit[100]{\upmu m}$ was left between fibers and SiPMs to prevent damage to the array surface.

The beam telescope offered a spatial resolution of approximately $\unit[30]{\upmu m}$ in the $x$-coordinate and $\unit[10]{\upmu m}$ in the $y$-coordinate. The fiber module prototype could be rotated around the $x$-axis in order to simulate different angles of incidence. 

For the readout of the SiPM arrays, electrical hybrid boards based on the IDEAS VA32/75 readout chip\cite{ideas} with 32 channels, a dynamic range of 36fC and a shaping time of $\unit[75]{ns}$ were developed. The gain of the SiPM arrays used is in the range of $10^6$, therefore the SiPM signals had to be attenuated by a factor of $150$ using a resistor network prior to sending them to the VA32 chip. The VA32/75 is compatible to the readout chip used for the AMS silicon tracker project so the same readout electronics required for the AMS silicon tracker was used to read out the optical hybrids\cite{alpat00}. 

\section{Test Results}

\begin{figure}[t]
\centering
\includegraphics[width=\columnwidth]{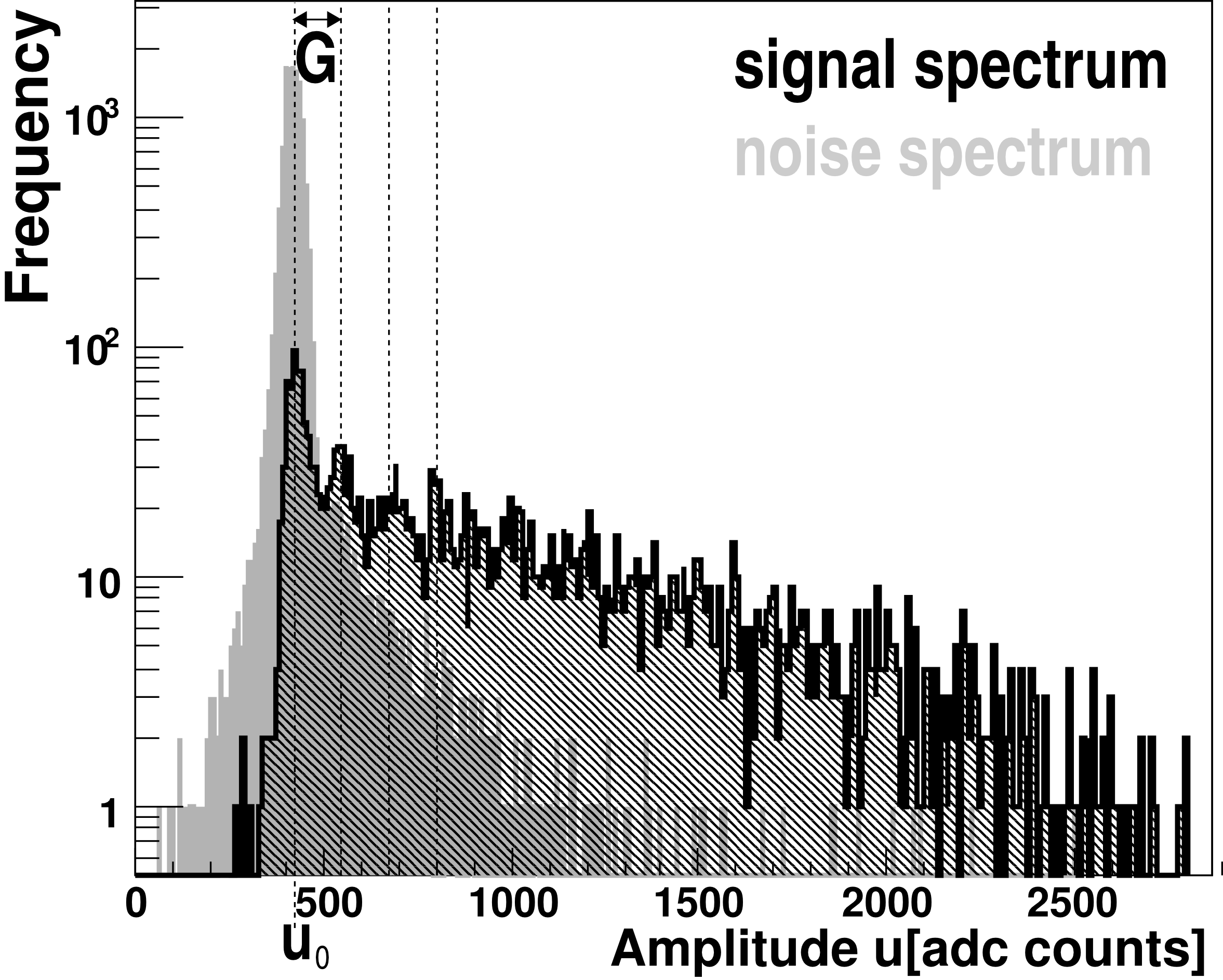}
\caption{The noise spectrum and the signal spectrum for particle tracks near the measured SiPM channel position for one channel of a MPPC 5883.}
\label{darksignal}
\end{figure}

Single-track events with isolated hits in both Si strip detectors are selected offering the particle position in $x$ and $y$ in the fiber module $(x_p, y_p)$. From each of the SiPM array channels we have the amplitude $u$ in ADC counts. The average position $y_f(x)$ of the fibers connected to one SiPM array channel with respect to the beam telescope is determined using the mean amplitude $\overline{u}(x_p, y_p)$ averaged over all events which were recorded for one configuration. The straight line $y_f(x) = a x + y_0$ that maximizes $\int_{-\infty}^{\infty}\mathrm{d}x~\overline{u}\left(x, a x + y_0\right)$ gives the most probable fiber position.  $y_f(x)$ is used to distinguish between the signal spectrum $S(u)$ and the pure noise spectrum $S_d(u)$ of the SiPM array (Fig.\ref{darksignal}). The signal spectrum which depends on the gain $G$ of the SiPM, the width $\sigma_0$ representing the electronic noise in the SiPM and readout chain, the variance of the gain $\sigma_G^2$, pedestal amplitude $u_0$ and the probability $p(n)$ to measure $n$ pixel discharges during one sampling is described by:
$$S(u) = \sum_{n=0}^{\infty} \frac{p(n)}{\sqrt{2 \pi} \sigma_n}\exp\left({-\frac{\left( u - u_n\right)^2}{\sigma_n^2}}\right)$$ where $u_n = u_0 + n \cdot G$ and $\sigma_n = \sqrt{\sigma_0^2 + n \cdot \sigma_G^2}$.
The gain $G$ and the pedestal position $u_0$ are determined from the signal spectrum. They are used to determine probability to see $n$ fired pixels for the dark spectrum $p_{\mbox{\tiny{d}}}(n)$. Assuming that the number of random noise discharges $n_{\mbox{\tiny{d}}}$ during one sampling follows the poisson distribution $\Pi(n)$, the mean number of noise pixel discharges $\overline{n}_d$  is determined from $\Pi(0) = p_{\mbox{\tiny{d}}}(0)$. The pixel crosstalk probability $\xi$ \cite{otte07} which leads to a deviation of $p_{\mbox{\tiny{d}}}(n)$ from poisson statistics for $n \ge 1$ is determined as $\xi = 1 - \frac{p_{\mbox{\tiny{d}}}(1)}{\Pi(1)}$. The MPPC 5883 had a much higher pixel crosstalk probability during the Beamtest at a lower DCR as shown in (Tab.\ref{sipmprops}).

\begin{table}[t]
\caption{Measured SiPM properties}
\begin{tabular*}{\columnwidth}{@{\extracolsep\fill}lrr}
\hline
Type & $ p_{\mbox{\tiny{d}}}(0)/\%$ & $\xi/\%$\\
\hline 
FBK-irst 2007 &  $93 \pm 2$ & $7 \pm 3$\\
MPPC 5883 & $96.8 \pm 0.6$ & $32 \pm 3$\\
\hline
\end{tabular*}
\label{sipmprops}
\end{table}

\subsection{Light Yield}

\begin{figure}[t]
\centering
\includegraphics[width=\columnwidth]{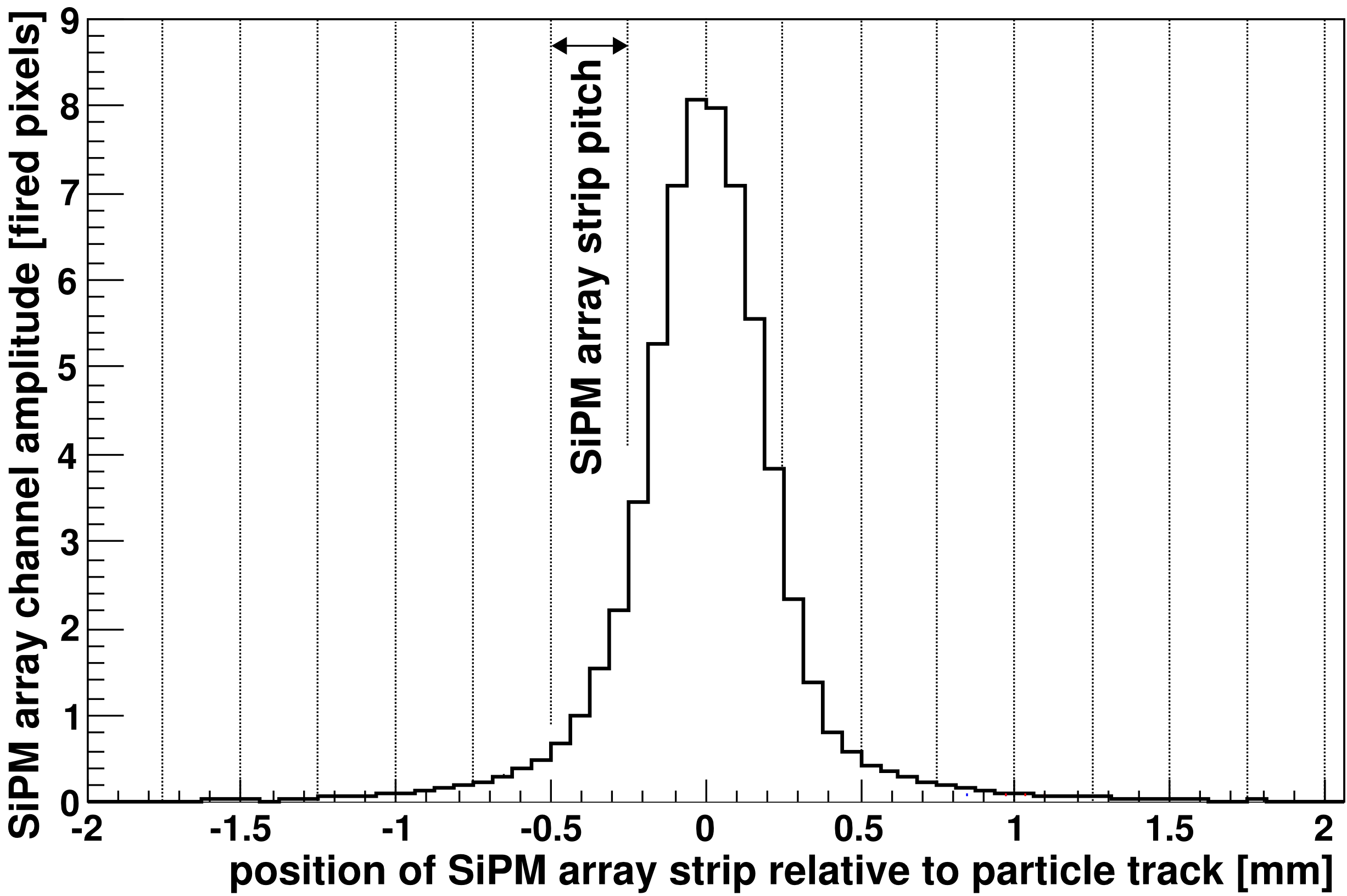}
\caption{The mean shape of a signal cluster for a single MPPC 5883.}
\label{clustershape}
\end{figure}

\begin{figure}[t]
\centering
\includegraphics[width=\columnwidth]{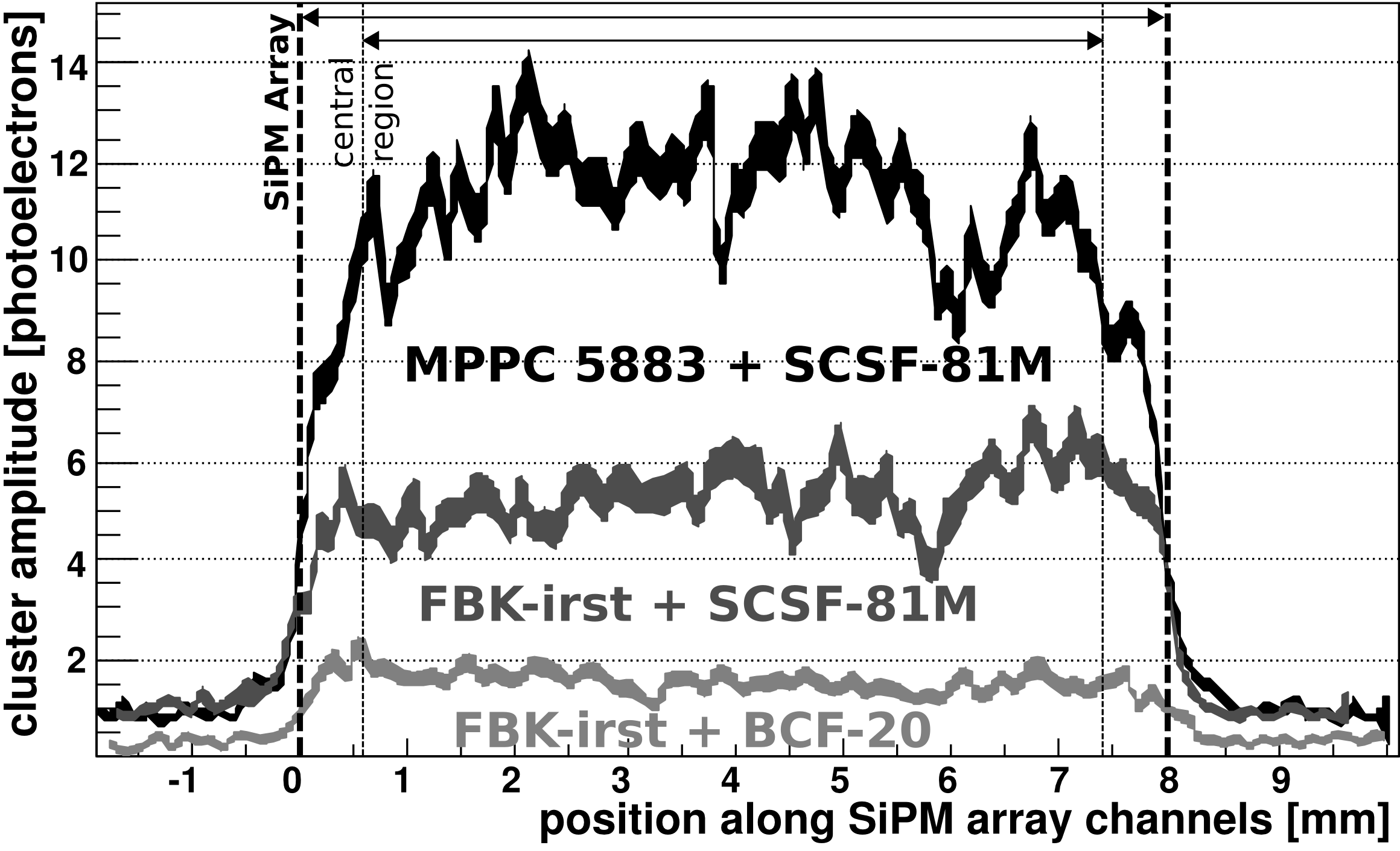}
\caption{The average cluster amplitudefor 3 different SiPM arrays from the beam test against the $y$-coordinate for normal particle incidence.}
\label{lightyield}
\end{figure}

The photon output of the scintillating fibers for a single particle track illuminates multiple SiPM array strips (Fig.\ref{clustershape}).  The distribution of photons on the SiPM array is smeared because the fiber pitch does not match the array readout pitch and because of the optical gap between SiPM array and fiber ($\unit[100]{\upmu m}$ for the FBK-irst 2007 and $\unit[275]{\upmu m}$ of optical glue for the MPPC 5883). Therefore signal clusters $C$ were detected for each event and array selecting the channel $i \in {1, .., 32}$ with the highest normalized amplitude $\alpha(i) = \frac{u(i) - u_0(i)}{G(i)}$ as a seed for the signal cluster. Next all channels $j$ with $\alpha(j) > 0.5$ are recursively added to $C$ if the neighbouring channel already belongs in $C$. To calculate the cluster amplitude $A(C)$ which equals the number of detected photons, the pixel crosstalk and the limit on the dynamic range of each SiPM array channel imposed by the limited number of pixels $N_{\mbox{\tiny{pix}}}$ has to be corrected for. Using the channel amplitude corrected for the limited dynamic range $\alpha_c(j) = N_{\mbox{\tiny{pix}}}\ln\left(1 - \frac{\alpha(j)}{N_{\mbox{\tiny{pix}}}}\right)$ and the correction term for pixel crosstalk $X = \frac{1 - \xi^{\lfloor\alpha_c(j)\rceil}}{1 - \xi}$ gives us $A(C) = \sum\limits_{j \in C} \frac{\alpha_c(j)}{X}$. The average number of photoelectrons for normal incidence is $9.6 \pm 1.2$ for the combination MPPC 5883 and SCSF-81M, $5.3 \pm 0.6$ for the combination FBK-irst 2007 and SCSF-81M and $1.9 \pm 0.2$ for BCF-20 fibers read out by the FBK-irst 2007 (Fig.\ref{lightyield}).

\subsection{Spatial Resolution}

\begin{figure}[t]
\centering
\includegraphics[width=\columnwidth]{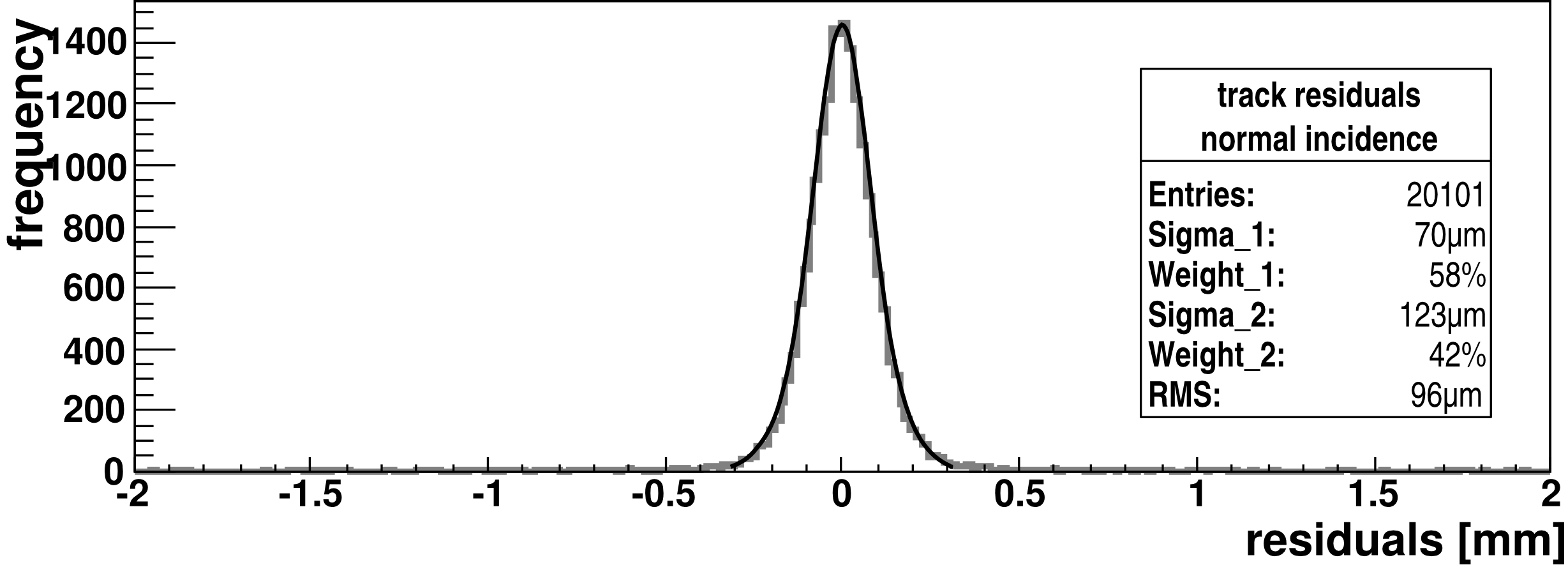}
\caption{The fitted track residuals for one MPPC 5883 reading out SCSF-81M fibers.}
\label{residuals}
\end{figure}

\begin{figure}[t]
\centering
\includegraphics[width=\columnwidth]{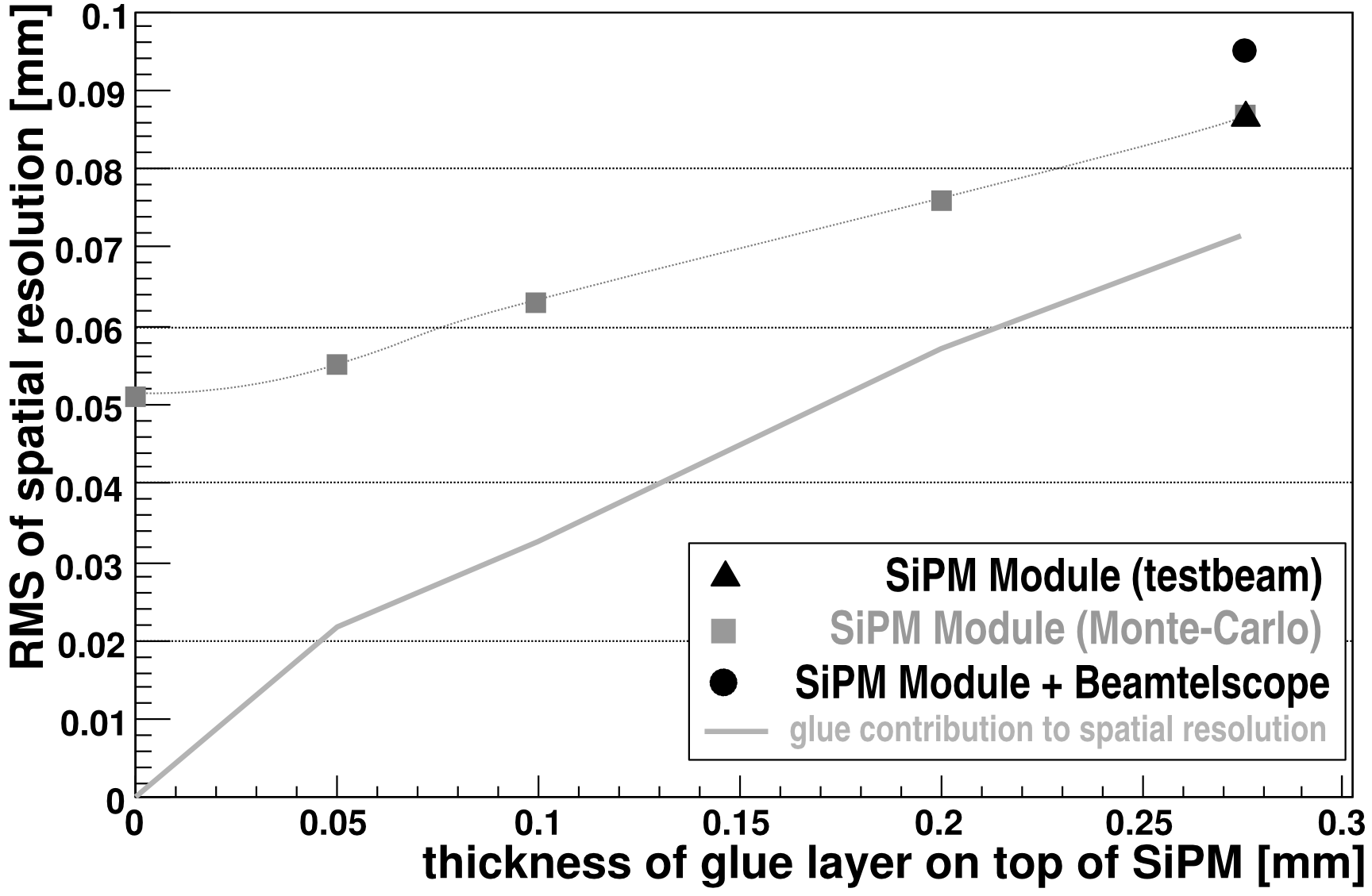}
\caption{The spatial resolution for the MPPC 5883/SCSF-81M prototype measured during the beamtest and the expected effect of reducing the thickness of the $\unit[0.275]{\upmu m}$ thick glue layer according to Monte-Carlo simulations.}
\label{spatialresprojection}
\end{figure}

The particle position from the fiber module $y_m$ is determined for all signal clusters $C$ with a cluster amplitude $A(C) > 0.5$, the channel amplitudes $\alpha_c(i)$ of the channels $i$, the measured average position $y_{f,i}(x)$ of the fibers connected to the channel $i$ and the $x$-coordinate of the particle $x_p$ from the beam telescope as: $y_m = \sum\limits_{i \in C} \frac{\alpha_c(i) y_{f,i}(x)}{A(C)}$. The residuals $\Delta_y = y_m - y_p$ (Fig.\ref{residuals}) are then fitted with the sum of two Gaussians $f(\Delta_y) = \frac{c_1}{\sqrt{2 \pi} \sigma_1}\exp\left(-{\frac{\Delta_y^2}{2 \sigma_1^2}}\right) + \frac{c_2}{\sqrt{2 \pi} \sigma_2}\exp\left(-{\frac{\Delta_y^2}{2 \sigma_2^2}}\right)$. The resulting rms of the distribution $f$ is the spatial resolution of $\rho = \sqrt{\frac{c_1 \sigma_1^2 + c_2 \sigma_2^2}{c_1 + c_2}}$. Based on a Monte-Carlo simulation, the contribution of the beam telescope and multiple scattering within the setup are calculated to be $\unit[38]{\upmu m}$ which is dominated by multiple scattering. The resulting spatial resolutions are $\unit[88]{\upmu m} \pm \unit[6]{\upmu m}$ at an efficiency of $\epsilon = 0.99$ for MPPC 5883/SCSF-81M and $\unit[110]{\upmu m} \pm \unit[16]{\upmu m}$ at an efficiency $\epsilon = 0.91$ for FBK-irst 2007/SCSF-81M subtracting the effect of the beam telescope. Tracks which are not passing in front of the central region of the SiPM array (Fig.\ref{lightyield}) are not considered for these results since in these cases part of the light signal is shared with neighbouring SiPM arrays.  

The gap between fiber and SiPM is suspected to play a role in optical crosstalk between SiPM array channels as well as smearing the position information carried by photons emitted by the scintillating fibers as has been studied by a dedicated Monte-Carlo simulation based on GEANT4\cite{geant4} (Fig.\ref{spatialresprojection}). These results indicate that a significant improvement in the spatial resolution is possible for the combination MPPC 5883/SCSF-81M if the thickness of the glue layer on top of the MPPC 5883 was reduced. 

\section{Conclusion}

The MPPC 5883 SiPM arrays and SCSF-81M fibers show the best overall performance among the tested components with a spatial resolution of $\unit[88]{\upmu m} \pm \unit[6]{\upmu m}$ and about $10$ detected photons for a MIP. The SCSF-81M fiber offers better mechanical properties as well as the higher light output by almost a factor 3 compared to the tested BCF-20 fiber. The primary cause of the worse spatial resolution and efficiency measured for the FBK-irst 2007  is the lower number of detected photons compared to the MPPC 5883. This effect is due nearly a factor two  lower geometrical fill factor of the FBK-irst arrays. A new array has been designed at FBK-irst in 2008 with two times higher filling factor, optimized for fiber tracking. The spatial resolution of the fiber module prototype can be improved by optimizing the coupling between SiPM and fiber.


\begin{thebibliography}{9}
\bibitem{gast07}
	H. Gast et al., NIM A, 581, pp. 151-155 (2007)
\bibitem{hamamatsu} 
	Hamamatsu Photonics K.K., Japan
\bibitem{piemonte06} 
	C. Piemonte, NIM A, 568, pp. 224-232 (2006)
\bibitem{piemonte07}
	C. Piemonte, R. Battiston et al., IEEE Trans. Nucl. Sci. 54, pp. 236-244 (2007)
\bibitem{greim08}
	R. Greim et. al., these proceedings
\bibitem{kuraray} 
	Kuraray CO., LTD., Japan
\bibitem{bicron} 
	Saint-Gobain Crystals, France
\bibitem{burger02}
	J. Alcaraz et al., NIM A 593, pp. 376-398 (2008)
\bibitem{ideas}
	IDEAS ASA, Norway
\bibitem{alpat00} 
	B. Alpat, NIM A, 439, pp. 53-64 (2000)
\bibitem{otte07}
	N. Otte, JINST TH 003 (2007)
\bibitem{geant4}
	S. Agostinelli et al., NIM A, 506, pp. 250-303 (2003)
\end{thebibliography}
\end{document}